\documentclass[sigplan,screen]{acmart}
\renewcommand\footnotetextcopyrightpermission[1]{}

\AtBeginDocument{%
  }

\setcopyright{acmlicensed}
\copyrightyear{2018}
\acmYear{2018}
\acmDOI{XXXXXXX.XXXXXXX}

\acmConference[Conference acronym 'XX]{Make sure to enter the correct
  conference title from your rights confirmation emai}{June 03--05,
  2018}{Woodstock, NY}
\acmISBN{978-1-4503-XXXX-X/18/06}

\acmSubmissionID{}

\usepackage{amsmath}
\usepackage{amsthm}
\usepackage{amssymb}
\usepackage{mathtools}
\usepackage{bm}
\usepackage{bbm}
\usepackage{booktabs}
\usepackage{xspace}
\usepackage{color,soul}
\usepackage{multirow}
\usepackage{multicol}
\usepackage{caption}
\usepackage{enumitem}

\usepackage{microtype}
\usepackage{graphicx}
\usepackage{subfigure}
\usepackage{booktabs}
\usepackage{hyperref}
\usepackage{nicefrac}
\usepackage[normalem]{ulem}
\usepackage{tikz}
\usetikzlibrary{trees}
\usepackage{float}
\usepackage{subfigure}
\usepackage[ruled,noend,linesnumbered,vlined]{algorithm2e}
\usepackage{algorithmic}
\usepackage{textcomp}
\usepackage{array}
\usepackage{amsmath}
\usepackage{adjustbox}

\usepackage{listings}
\definecolor{myred}{rgb}{1.0,0.7,0.8}
\definecolor{mygreen}{RGB}{0,166,0}
\definecolor{lightgreen}{rgb}{0.56, 0.93, 0.56}
\definecolor{myorange}{RGB}{252,107,4}
\definecolor{darkgreen}{RGB}{0,153,102}
\definecolor{lightblue}{rgb}{0.53, 0.81, 0.92}
\definecolor{lightgray}{gray}{0.9}

\newcommand{\mysubsubsection}[1]{{\textbf{\textit{#1}.\xspace}}}

\DeclareMathOperator*{\argmin}{arg\,min}

\newcommand{\system}{{\texttt{ByteScale}}\xspace}

\newtheorem*{assumption*}{\assumptionnumber}
\providecommand{\assumptionnumber}{}


\begin{document}

\title{ByteScale: Efficient Scaling of LLM Training with a 2048K Context Length on More Than 12,000 GPUs}

\author{Hao Ge}
\authornote{Equal Contribution.}
\email{gehao@stu.pku.edu.cn}
\affiliation{
\institution{Peking University}
\country{}
}

\author{Junda Feng}
\authornotemark[1]
\email{fengjunda.aml@bytedance.com}
\affiliation{
\institution{ByteDance Seed}
\country{}
}

\author{Qi Huang}
\authornotemark[1]
\email{huangqi.lucky@bytedance.com}
\affiliation{
\institution{ByteDance Seed}
\country{}
}

\author{Fangcheng Fu}
\authornote{Corresponding Authors.}
\email{ccchengff@pku.edu.cn}
\affiliation{
\institution{Peking University}
\country{}
}

\author{Xiaonan Nie}
\email{niexiaonan@bytedance.com}
\affiliation{
\institution{ByteDance Seed}
\country{}
}

\author{Lei Zuo}
\email{zuo.lei@bytedance.com}
\affiliation{
\institution{ByteDance Seed}
\country{}
}

\author{Haibin Lin}
\authornotemark[2]
\email{haibin.lin@bytedance.com}
\affiliation{
\institution{ByteDance Seed}
\country{}
}

\author{Bin Cui}
\authornotemark[2]
\email{bin.cui@pku.edu.cn}
\affiliation{
\institution{Peking University}
\country{}
}

\author{Xin Liu}
\authornotemark[2]
\email{liuxin.ai@bytedance.com}
\affiliation{
\institution{ByteDance Seed}
\country{}
}

%% article.
\begin{abstract}
Scaling long-context ability is essential for Large Language Models (LLMs). To amortize the memory consumption across multiple devices in long-context training, inter-data partitioning (a.k.a. Data Parallelism) and intra-data partitioning (a.k.a. Context Parallelism) are commonly used. Current training frameworks predominantly treat the two techniques as orthogonal, and establish static communication groups to organize the devices as a static mesh (e.g., a 2D mesh). However, the sequences for LLM training typically vary in lengths, no matter for texts, multi-modalities or reinforcement learning. The mismatch between data heterogeneity and static mesh causes redundant communication and imbalanced computation, degrading the training efficiency.

In this work, we introduce \system, an efficient, flexible, and scalable LLM training framework for large-scale mixed training of long and short sequences. The core of \system is a novel parallelism strategy, namely Hybrid Data Parallelism (HDP), which unifies the inter- and intra-data partitioning with a dynamic mesh design. In particular, we build a communication optimizer, which eliminates the redundant communication for short sequences by data-aware sharding and dynamic communication, and further compresses the communication cost for long sequences by selective offloading. Besides, we also develop a balance scheduler to mitigate the imbalanced computation by parallelism-aware data assignment. We evaluate \system with the model sizes ranging from 7B to 141B, context lengths from 256K to 2048K, on a production cluster with more than 12,000 GPUs. Experiment results show that \system outperforms the state-of-the-art training system by up to $7.89\times$.
\end{abstract}

\settopmatter{printacmref=false}
\maketitle
\pagestyle{plain}

\vspace{-5pt}
\section{Introduction}
\label{sec:intro}
In recent years, large language models (LLMs) have achieved remarkable success across various domains. The impressive performance of LLMs is attributed to increased model sizes, larger volumes of training data, and longer context windows, all in accordance with the scaling law~\cite{scalinglaws}. The demand for long-context capabilities of LLMs has increased rapidly, as modern LLM applications like documents summarization~\cite{document_summary}, video understanding~\cite{video_agent,longvlm}, agent interaction~\cite{agent_interactive} and code completion~\cite{code_understanding}, require the model to understand long-range dependencies. It has driven many organizations to extend their models' context lengths. For instance, Meta's LLaMA3~\cite{llama3} and OpenAI's GPT-4o~\cite{gpt4o} support 128K contexts, Anthropic's Claude3~\cite{claude3} supports 200K, and Google's Gemini-1.5 Pro~\cite{gemini2m} supports up to 2M contexts.

A fundamental challenge in scaling to a long context is the quadratic scaling of memory and computation for self-attention. Flash Attention~\cite{flash_attn,flash_attn_v2} has been proposed to reduce the memory complexity from $O(S^2)$ to $O(S)$, where $S$ is the sequence length. To further scale the context length, it's necessary to partition the sequences across multiple devices.
There are broadly two categories: inter-data partitioning (a.k.a. Data Parallelism, DP~\cite{pytorch_ddp,horovod,dist_belief}) distributes different sequences across the devices, while intra-data partitioning (a.k.a. Context Parallelism, CP~\cite{lightseq,ring_attn,striped_attn,context_parallel}) scatter a single sequence. 
Both categories evenly reduce the memory consumption on each device, while inevitably incurring extra communication overhead. 
Existing LLM training frameworks, such as Megatron-LM~\cite{megatron_1,megatron_2,megatron_3}, DeepSpeed~\cite{deepspeed,deepspeed_ulysses} and MegaScale~\cite{megascale}, treat the two categories as individual parallelism strategies, and establish DP$\times$CP communication groups to organize the devices as a static mesh (e.g., a 2D mesh), where the size of each CP group is dependent on the maximum sequence length (i.e., context length). 
Undoubtedly, it requires the sequences to be of the same length so that the training workloads across devices are uniform. 

Nevertheless, the sequences for LLM training usually vary in lengths. For one thing, sequence lengths typically exhibit skewed distribution in real-world datasets, no matter the text or multi-modal data. For another thing, inference-time scaling (e.g. OpenAI's o1~\cite{gpto1}, DeepSeek-R1~\cite{deepseek_r1}) increases the length of the Chain-of-Thought reasoning process, further exacerbates length heterogeneity for reinforcement learning. 
When facing the sequences with variable lengths, existing frameworks can only configure the size of CP groups to be large enough to handle the longest sequences (yielding a small DP size), and each sample needs to be evenly partitioned across the entire CP group, regardless of sequence length, degrading the overall training efficiency. 

In particular, the mismatch between data heterogeneity and static system design causes two main challenges (detailed in \S\ref{sec:motivation}). \textcircled{1} \textbf{Redundant Communication}: It is common practice to pack~\cite{seq_packing} shorter sequences into a single one up to the context length and configure a sufficient CP size to prevent out-of-memory (OOM) errors. However, all short sequences have to undergo the same partitioning and communication process as long sequences, even if it is unnecessary. Worse yet, CP requires $O(S^2)$ computation to overlap $O(S)$ communication, which is challenging for short sequences. \textcircled{2} \textbf{Imbalanced Computation}: Although tokens are evenly partitioned across devices by CP and memory is balanced, execution times still vary. This is because the computational complexity of each token is related to the original sequence length, which is $O(S^2)$. The imbalanced computation causes some devices to fall into idle time for synchronization.

\textbf{Summary of Contributions.} To address the aforementioned challenges, we propose \system, an efficient, flexible, and scalable training framework designed for large-scale mixed training of long and short sequences. The main contributions are as follows:

\mysubsubsection{C1: Proposal of Hybrid Data Parallelism}
We propose a novel parallelism strategy, namely Hybrid Data Parallelism (HDP), which unifies both inter-data (DP) and intra-data partitioning (CP), and is defined to evenly distributing \underline{\textit{tokens}} across devices. It utilizes devices in the range of [1, DP$\times$CP] to flexibly process variable-length sequences.

\mysubsubsection{C2: Communication Optimizations}
To eliminate redundant communication for short sequences, HDP provides the ability of data-aware sharding, where dynamic communication groups are automatically built and each sequence will be processed with a minimal number of devices individually. Besides, HDP also provides selective offloading to further compress the communication cost for long sequences.

\mysubsubsection{C3: Balance Strategy}
To mitigate the imbalanced computation, we design a heuristic algorithm that reorganizes data assignment based on the characteristics of data and pipeline parallelism. Furthermore, for those devices with shorter execution times, we assign more micro batches, rather than the same number under the static system design.

\mysubsubsection{C4: Evaluation}
We conduct experiments on a production cluster with more than 12,000 GPUs, scaling the model size from 7B to 141B, and context length from 256K to 2048K. The results demonstrate that \system achieves up to $7.89\times$ of speedup compared to existing training approaches.

\vspace{-5pt}
% \vspace{-15pt}
\section{Background}
\label{sec:background}

\begin{figure}
\centering
\includegraphics[width=\linewidth]{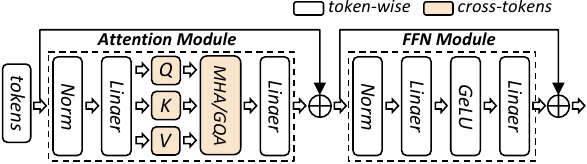}
\vspace{-20pt}
\caption{\small{the Architecture of Transformer layer}}
\label{fig:transformer_architecture}
\vspace{-15pt}
\end{figure}
% \vspace{-10pt}

\begin{figure}
\centering
\includegraphics[width=\linewidth]{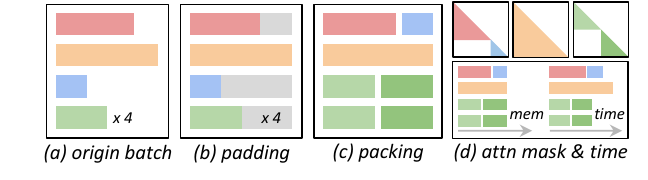}
\vspace{-20pt}
\caption{\small{Sequence Padding and Packing}}
\label{fig:pad_and_pack}
\vspace{-15pt}
\end{figure}

\begin{figure*}[t]
\centering
\includegraphics[width=\linewidth]{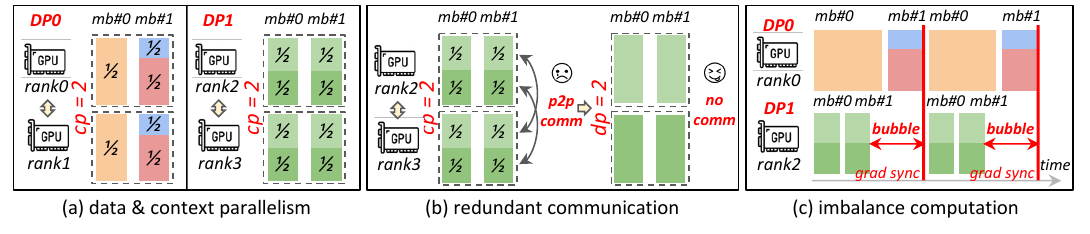}
\vspace{-20pt}
\caption{\small{Context Parallelism with Packing}}
\label{fig:cp_and_motivation}
\vspace{-10pt}
\end{figure*}
% \vspace{-5pt}

\begin{figure}
\centering
\includegraphics[width=\linewidth]{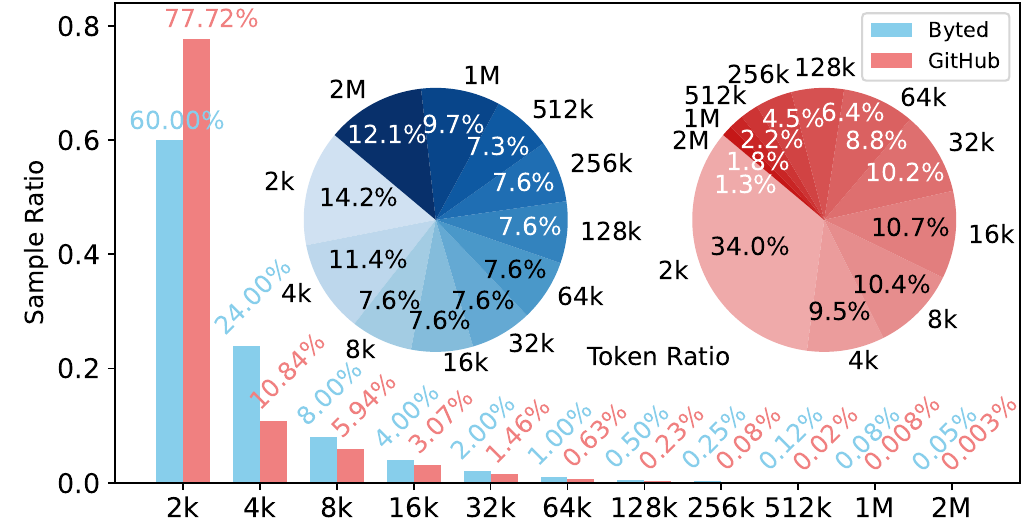}
\vspace{-20pt}
\caption{\small{Sample and Token Distribution in two Datasets}} % 4M tokens
\label{fig:sample_token_ratio}
\vspace{-10pt}
\end{figure}

\subsection{Transformers and Large Language Models}
\label{sec:background_large_language_models}
The transformer architecture~\cite{attn} has become the most popular and widely used foundational architecture for large language models (LLMs)~\cite{gpt3,gpt4,llama2,gemini} nowadays. It typically consists of a series of transformer layers, each comprising an attention module and a feed-forward network (FFN) module.

As shown in Figure~\ref{fig:transformer_architecture}, self-attention captures contextual information throughout the entire text, necessitating all tokens in the full sequence to participate in computation. In contrast, other operations like normalization, linear projection, and activation functions perform token-wise computations, allowing each token to be processed independently. 

% \vspace{-15pt}
\subsection{Distributed LLM Training}
\label{sec:background_distributed_llm_training}
As model sizes and training data continue to scale, distributed training techniques are indispensable in LLM training.

\mysubsubsection{Data Parallelism}
Data parallelism (DP)~\cite{pytorch_ddp,horovod,dist_belief} distributes the training data evenly across devices, while each device holds a replica of the model. During each training step, devices process their local data individually, and synchronize gradients globally to update the model. ZeRO series~\cite{zero} methods further enhance the scalability of DP.

\mysubsubsection{Model Parallelism}
Model parallelism distributes the model across devices, including tensor parallelism (TP)~\cite{megatron_1} and pipeline parallelism (PP)~\cite{gpipe,pipedream,pipedream_flush}. TP performs intra-operation partitioning, dividing operations and parameters within a layer across devices (e.g. Row- and Col-Parallel Linear in Megatron-LM~\cite{megatron_1}). It requires communication of intermediate results (activations), and is typically used within a single node. PP employs inter-operation partitioning, segmenting the model layers into different stages. It requires only the exchange of activations between consecutive stages via peer-to-peer (P2P) communication, enabling model partitioning across multiple nodes.

\mysubsubsection{Hybrid Parallelism}
Hybrid parallelism combines various parallel strategies to enhance training efficiency. Particularly, Megatron-LM employs the 3D parallelism strategy~\cite{megatron_1,megatron_2,megatron_3} by integrating DP, TP, and PP, making it a mainstream approach for large-scale model training today. 

\mysubsubsection{Gradient Accumulation}
To improve the efficiency and convergence, LLMs typically require large batch size~\cite{chinchilla,palm,llama2} (e.g. it is common practice to apply nearly 30\textasciitilde80M tokens per batch for LLM training in the cluster with 10K GPUs). Constrained by hardware memory, processing the entire large batch at once is infeasible. Gradient accumulation divides each global batch (i.e., the sampled data in each training step) into multiple micro-batches. The gradients from these micro-batches are accumulated to equal the gradient as if the entire global batch were processed in a single pass.

\subsection{Padding and Packing}
\label{sec:background_padding_and_packing}
To support variable-length sequences in current static parallelism strategies, techniques such as padding and packing are necessary. As illustrated in Figure~\ref{fig:pad_and_pack}, padding pads the sequences in the same batch to be of the same length, but causes wasted computation. Packing~\cite{seq_packing} concatenates multiple sequences into a single one without padded tokens. It employs a special segmented attention mask to ensure that each sequence is processed independently by self-attention.
% \vspace{-5pt}

\subsection{Long Context Training}
\label{sec:background_long_context_training}
As self-attention exhibits both time and memory complexity of $O(S^2)$, when the context length scales, this quadratic complexity becomes a bottleneck. Flash Attention~\cite{flash_attn,flash_attn_v2} optimizes memory I/O and employs the tiling technique to reduce memory complexity from $O(S^2)$ to $O(S)$, while still maintaining $O(S^2)$ time complexity. Context Parallelism (CP)~\cite{lightseq,ring_attn,striped_attn,context_parallel} further partitions the sequence across $N$ devices, reducing the memory from $O(S)$ to $O(\frac{S}{N})$. Following Figure~\ref{fig:transformer_architecture}, CP shards QKV along the sequence dimension, and cross-tokens operations require KV slices to be exchanged across devices using a ring-style P2P communication, which overlaps with computation. This technique is also applicable to packed sequences, and we will detail its implementation in §\ref{sec:method_implementation}. Notably, each subsequence must also be sharded across all CP ranks, as illustrated in Figure~\ref{fig:pad_and_pack}(c) and ~\ref{fig:cp_and_motivation}(a).
\section{Obervation \& Motivation}
\label{sec:motivation}

\begin{figure*}[t]
\centering
\includegraphics[width=\linewidth]{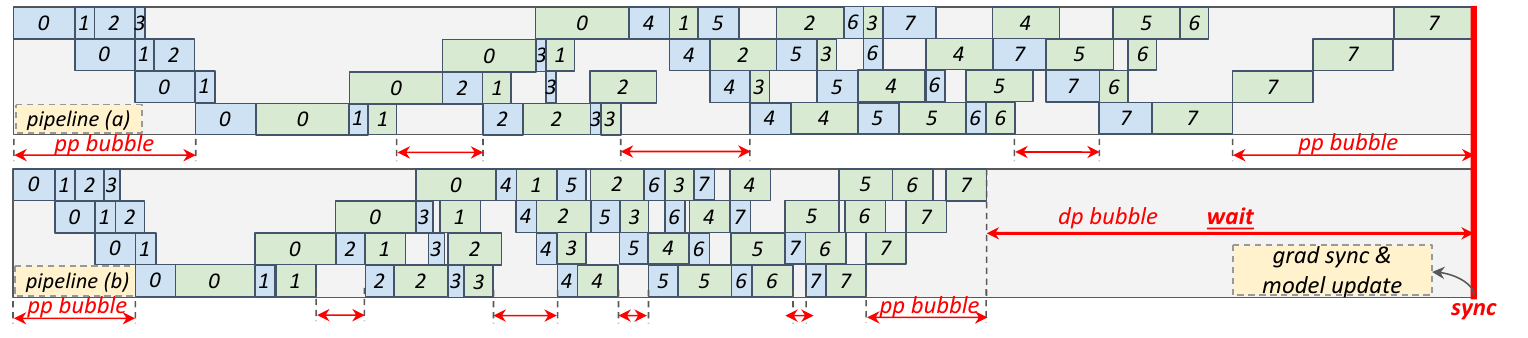}
\vspace{-20pt}
\caption{\small{Imbalanced Data and Pipeline Parallelism}}
\label{fig:dp_pp_imbalance}
\vspace{-10pt}
\end{figure*}
% \vspace{-10pt}

\begin{figure}
\centering
\includegraphics[width=\linewidth]{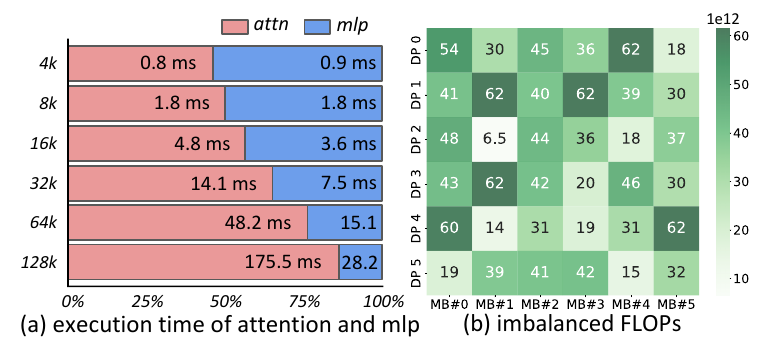}
\vspace{-20pt}
\caption{\small{Imbalanced Computation}}
\label{fig:imbalanced_flops}
\vspace{-10pt}
\end{figure}

\subsection{Data Heterogeneity}
\label{sec:motivation_data_heterogeneity}

LLMs are trained on sequences data. As mentioned in \S\ref{sec:intro}, the training data typically consists of variable-length sequences.
There exist two observations and one significant challenge: 

\mysubsubsection{Observation 1: sequence lengths exhibit skewed distribution in real-world datasets}
As shown in Figure ~\ref{fig:sample_token_ratio}, we profiled the sample and token distribution of two datasets: an open-source dataset \textit{GitHub} and a productive dataset \textit{Byted} for long-context training. We observed that both of them exhibit a skewed distribution in sequence lengths. For instance, in the \textit{Byted} dataset, if we randomly sample a global batch, nearly 80\% of the samples are 4K tokens or shorter, while only 0.05\% of the samples can reach 2M tokens. However, from the perspective of token distribution, those 0.05\% of the samples (>=2M) contribute 12.1\% of the tokens in the global batch, and 1\% of the samples (>=128K) contribute 44.3\%. Although the \textit{GitHub} dataset has a lower proportion of long sequences, 16.2\% of its tokens come from sequences exceeding 128K, demonstrating significant data heterogeneity. 

\mysubsubsection{Observation 2: mixing long and short sequences enhances model performance} The existing work~\cite{long_short_mix} has demonstrated that training exclusively on long-context data can lead to a decline in short-context performance. LLaMA3 report~\cite{llama3} indicates that when training a model with 128K context, mixing 0.1\% of long data with the original short data optimizes the performance across both short-context and long-context benchmarks. DeepSeek-R1~\cite{deepseek_r1} presents the average response length on the training set during the RL process, demonstrating that gradually increasing and diverse response lengths help improve model performance.

\mysubsubsection{Challenge: data heterogeneity leads to efficiency degradation} 
Although mixed training of long and short sequences is common and beneficial for model performance, it introduces new challenges. The static parallelism strategies used in existing systems are not well-suited to handle dynamic workloads. This causes issues of redundant communication (§\ref{sec:motivation_redundant_communication}) and imbalanced computation (§\ref{sec:motivation_imbalanced_computation}), which we will discuss in more detail below.

\subsection{Redundant Communication}
\label{sec:motivation_redundant_communication}
Existing systems apply static parallelism strategies throughout the training process. Typically, they assume that all (packed) sequences are of the same length and set a fixed CP degree to amortize them across enough devices, thereby avoiding OOM errors. As mentioned in §\ref{sec:background_padding_and_packing}, to handle variable-length sequences, it is common to pack sequences up to the context length. However, as depicted in Figure \ref{fig:cp_and_motivation}(a)-(b), all sequences have to be partitioned across the entire CP group, even if it is unnecessary for shorter ones.

For instance, assuming that each device has a capacity of 8K tokens, to train an LLM with a context length of 1M tokens, a CP degree of 128 is required. This configuration necessitates 128 individual devices to process a sequence of 1M tokens. Concurrently, a large number of shorter sequences, such as those with lengths of 4K, 8K, and 16K tokens, are packed up to 1M tokens and processed in a CP group with 128 devices. As depicted in Figure~\ref{fig:dist_attn_with_packing}, each subsequence within the packed sequence needs to be partitioned into 128 chunks across CP ranks, performing ring-P2P communication. In fact, it is unnecessary to perform cross-device partitioning and communication for sequences with lengths under 8K. For those sequences with 16K tokens, only two CP ranks are required. Using the same CP degree as for the maximum sequence length leads to excessive redundant communication for these shorter sequences. This issue is exacerbated when sequence lengths are highly skewed.

\begin{figure}
\centering
\includegraphics[width=\linewidth]{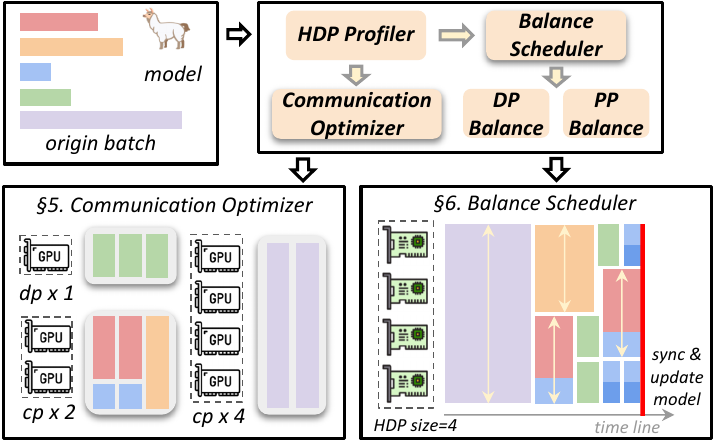}
\vspace{-20pt}
\caption{\small{\system Overview}}
\label{fig:overview}
\vspace{-10pt}
\end{figure}

\begin{figure*}
\centering
\includegraphics[width=\linewidth]{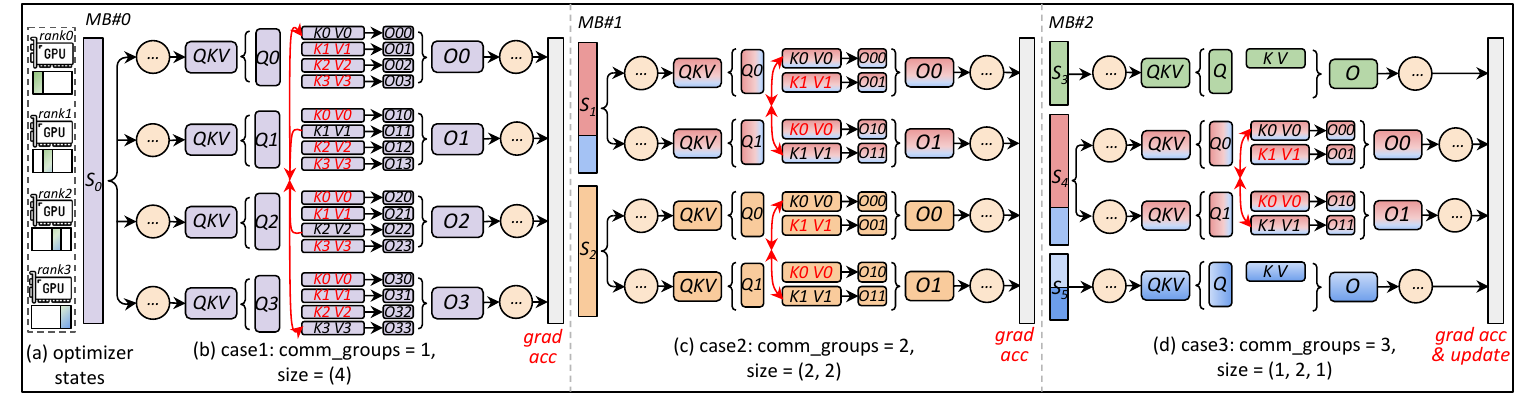}
\vspace{-20pt}
\caption{\small{Illustration of HDP}}
\label{fig:hdp}
\vspace{-10pt}
\end{figure*}

\subsection{Imbalanced Computation}
\label{sec:motivation_imbalanced_computation}

\mysubsubsection{Imbalanced FLOPs}
Although Flash Attention enables linear packing with $O(S)$ memory complexity, the computational complexity for each subsequence remains $O(S^2)$. As depicted in Figures~\ref{fig:pad_and_pack}(d) and ~\ref{fig:cp_and_motivation}(c), even if two packed sequences contain the same number of tokens, their actual computational workloads differ, which are proportional to the areas of attention mask. As shown in Figure~\ref{fig:imbalanced_flops}(a), when the context length is shorter than 8K tokens, the $O(S^2)$ term is relatively insignificant, allowing packing to effectively balance workloads for both memory and computation. However, for long-context training tasks, the $O(S^2)$ term becomes the predominant component of the computation, leading to significant time imbalances across different packed sequences.

To provide an intuitive explanation, we sampled a global batch of 1.2M tokens from the \textit{GitHub} dataset and randomly packed them into micro-batches of up to 32K tokens, aligning with the model's context length. As shown in Figure~\ref{fig:imbalanced_flops}(b), we recorded the FLOPs (Floating Point Operations) for each micro-batch and observed significant variability, indicating that the execution time for each micro-batch also differs.

\mysubsubsection{Imbalanced Data and Pipeline Parallelism}
The imbalanced execution times across micro-batches further degrade the efficiency of data and pipeline parallelism. In data parallelism, all DP ranks must execute the same number of micro-batches, and then synchronize gradients before the model update. As illustrated in Figure~\ref{fig:cp_and_motivation}(c), rank-2 processes tokens with fewer FLOPs than rank-0, leading to idle time (i.e. DP Bubble) as it waits for synchronization. In pipeline parallelism, there are two types of ``bubbles'': the PP bubble occurs within a single pipeline, and the DP bubble occurs across different pipelines (different DP groups). Aside from PP bubbles during the warmup and cooldown phases, imbalanced FLOPs between micro-batches prevent the execution time on different devices from overlapping as they would in an ideal pipeline. This leads to extra PP bubbles caused by inter-stage waiting, as shown in Figure ~\ref{fig:dp_pp_imbalance}. Additionally, since each micro-batch is executed sequentially across $d_{\text{pp}}$ different stages in the pipeline, any DP bubble will be magnified by a factor of $d_{\text{pp}}$. For example, consider two pipelines illustrated in Figure~\ref{fig:dp_pp_imbalance}, the micro-batches 0 and 7 in the pipeline (a) have a longer forward and backward execution time compared to those in the pipeline (b). Under $d_{\text{pp}}=4$, this time gap is magnified fourfold. Consequently, after executing 8 micro-batches, the pipeline (b) falls into a prolonged idle period, waiting for gradient synchronization. This causes the DP bubble to account for over 30\% of the total execution time, far exceeding the normal pipeline bubble time.

\section{\system Overview}
\label{sec:method_overview}

We present \system to address these challenges. As shown in Figure~\ref{fig:overview}, it consists of three main components. \underline{Profiler} is to profile the environment, model configuration, data distribution, and build cost models for other components. \underline{Communication Optimizer} is to improve the communication efficiency for both short and long sequences by data-aware sharding, dynamic communication, and selective offloading. \underline{Balance Scheduler} is to solve the imbalanced computation by parallelism-aware data assignment.

\section{Communication Optimizer}
\label{sec:method_comminication_optimizer}
This section describes how \system optimizes communication overhead. First, it reduces redundant communication for short sequences by dynamic sequence sharding and communication. Second, it further compresses the communication cost for long sequences by selective offloading. 

\subsection{Data-Aware Sharding and Communication}\label{sec:method_data_aware_sharding_and_communication}
\mysubsubsection{Hybrid Data Parallelism}
To begin with, we introduce a novel parallelism strategy, namely Hybrid Data Parallelism (HDP), to enable efficient training for different levels of sequence lengths. Both DP and CP partition training data across devices. DP performs inter-data partitioning by distributing different samples evenly across devices, while CP performs intra-data partitioning by sharding a single sample across devices. HDP unifies both inter-data and intra-data partitioning and is defined to evenly distribute \underline{\textit{tokens}} across devices. It can replace traditional DP and CP, with the parallel degree of HDP equivalent to the product of the degrees of DP and CP (i.e. $d_{\text{hdp}} = d_{\text{dp}} \times d_{\text{cp}}$). 

Unlike DP and CP, which require all DP/CP ranks to perform consistent behavior in computation or communication (e.g. CP requires all CP ranks to participate in homogeneous ring-P2P communication), HDP allows for heterogeneous behavior among HDP ranks. It has two key characteristics:
\begin{itemize}[label=$\triangleright$,leftmargin=*]
    \item[\textcircled{1}] \textbf{More Flexible Communication}: HDP only requires that different HDP ranks handle an equal number of tokens. This means that some HDP ranks may be assigned complete sequences (short sequences), as illustrated by $S_3$ and $S_5$ in Figure~\ref{fig:hdp}(d), while some other ranks may only handle the partial slice of a sequence (long sequences), as shown with $S_4$ in Figure~\ref{fig:hdp}(d). This necessitates establishing more flexible communication groups. For instance, in Figure~\ref{fig:hdp}(d), a communication group of size 2 is created only between rank-[1\textasciitilde2] to compute the distributed attention for $S_4$, while rank-0 and 3 can perform local computation without cross-device communication. In Figure~\ref{fig:hdp}(b), sequence $S_0$ is sharded into four slices, and a communication group of size 4 is created among rank-[0\textasciitilde3]. 
    
    \item[\textcircled{2}] \textbf{More Finer-Grained Communication}: Static parallel strategies require that the product of the parallel degrees equals the number of devices in the cluster, i.e. $d_{\text{dp}} \times d_{\text{cp}} \times d_{\text{tp}} \times d_{\text{pp}} = N_{\text{cluster}}$, where $d_{\text{tp}}$ and $d_{\text{pp}}$ are actually fixed based on model size. To utilize all the devices and maintain this divisibility, $d_{\text{dp}}$ and $d_{\text{cp}}$ can only be scaled by a limited factor, resulting in coarse granularity (e.g. assume each rank can handle 8K tokens, 512K can use <$d_{\text{dp}}=2$, $d_{\text{cp}}=64$>, while 768K needs $d_{\text{cp}}=96$ but must use <$d_{\text{dp}}=1$, $d_{\text{cp}}=128$>). Meanwhile, HDP can use any amount of ranks in $[1, d_{\text{hdp}}]$ to handle a sequence without considering the divisibility constraints (e.g. with $d_{\text{hdp}} = d_{\text{dp}} \times d_{\text{cp}} = 128$, HDP can use 96 ranks to handle a 768K sequence while use rest 32 ranks to handle 32 $\times$ 8K sequences individually). % with finer granularity.
\end{itemize}

\begin{figure}
\centering
\includegraphics[width=\linewidth]{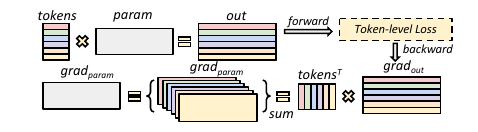}
\vspace{-20pt}
\caption{\small{Token-Level Gradient}}
\label{fig:token_grad}
\vspace{-10pt}
\end{figure}

\begin{figure}
\centering
\includegraphics[width=\linewidth]{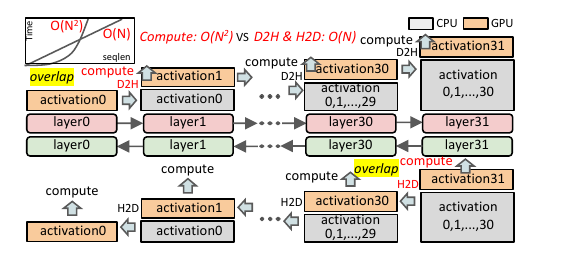}
\vspace{-20pt}
\caption{\small{Per-Layer Activation Offloading}}
\label{fig:per_layer_offload}
\vspace{-10pt}
\end{figure}

\mysubsubsection{NCCL Buffer Optimization}
Creating NCCL communication groups incurs extra overhead. Firstly, the process of establishing a communication group is inherently slow, and dynamically creating new groups for each sequence can significantly reduce training efficiency. Secondly, creating an excessive number of communication groups can consume an additional 5\textasciitilde10GB of memory per GPU for NCCL buffers, further reducing the available memory. Fortunately, distributed attention utilizes P2P communication. With a global communication group across all HDP ranks, P2P communication between any two devices can directly reuse the existing group, thereby alleviating the time and memory pressure associated with creating temporary communication groups.

\mysubsubsection{Optimizer States Sharding}
HDP evenly partitions tokens across devices, and will shard neither model parameters nor gradients. This means that HDP ranks replicate the model states like DP. Consequently, the ZeRO series technique is also suitable to HDP, as shown in Figure~\ref{fig:hdp}(a), HDP utilizes ZeRO-1 across all the HDP ranks to maximally shards the optimizer states, minimizing the memory usage.

\mysubsubsection{Loss and Model Update}
Even though HDP ranks may perform different heterogeneous communications across different micro-batches, the final gradient for a parameter is equivalent to that obtained in standard DP. As shown in Figure~\ref{fig:token_grad}, each token contributes a gradient to the parameter $\theta_n$, and the final gradient, denoted as $G_{\theta_n}$, is the sum over gradients from all tokens in global batch (denoted as $\mathbb{B}$). Let $\text{grad}(j, \theta_n)$ represent the gradient from the token $j$ to the parameter $\theta_n$. Then $G_{\theta_n}$ can be presented as: 

\begin{equation}\label{eq:grad_sum_dp}
\small
G_{\theta_n} = \sum\nolimits_{S_i \in \mathbb{B}} \left( \sum\nolimits_{j \in S_i} \text{grad}(j, \theta_n) \right)
\end{equation}

Since parameters are replicated and tokens are evenly distributed across HDP ranks (denoted as $\mathbb{R}$), the local accumulated gradient corresponds to the partial sum of gradients from tokens assigned to each rank (denoted as $\mathbb{B}^r$, i.e. micro-batches in rank $r$). Consequently, similar to DP, a global collective communication like All-Reduce or Reduce-Scatter will be performed across all HDP ranks to aggregate partial gradients. This also yields the gradient $G_{\theta_n}$ from all tokens:

\begin{equation}\label{eq:grad_sum_hdp}
\small
G_{\theta_n} = \sum\nolimits_{r \in \mathbb{R},\ \mathbb{B}^r \in \mathbb{B}} \left( \sum\nolimits_{m \in \mathbb{B}^r} \left( \sum\nolimits_{j \in m} \text{grad}(j, \theta_n) \right) \right)
\end{equation}

The Eq.(\ref{eq:grad_sum_hdp}) is equivalent to Eq.(\ref{eq:grad_sum_dp}), and ensures that the result of gradient accumulation in HDP is equivalent to that in standard DP. Moreover, since we calculate the gradient $G_{\theta_n}$ over all tokens in the global batch, it also needs to be scaled by the total amount of tokens, as we implement this by the \textit{token-level loss}, which scales the loss by the token amount rather than sample amount.

\subsection{Data-Aware Selective Offloading}

\begin{figure}
\centering
\includegraphics[width=\linewidth]{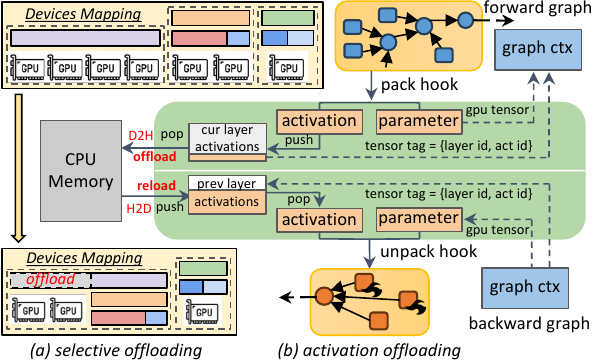}
\vspace{-20pt}
\caption{\small{Data-Aware Selective Offloading}}
\label{fig:offload}
\vspace{-10pt}
\end{figure}
% \vspace{-4pt}

\mysubsubsection{Activation Offloading}
The activation size is proportional to the sequence length. Constrained by GPU memory, longer sequences require more HDP ranks to distribute the activation. For example, processing a sequence with 1M tokens requires 128 ranks if each rank can handle 8K tokens, which is usually unaffordable with today's expensive GPU resources. In practice, modern GPU servers are typically equipped with CPU memory that far exceeds GPU memory. Therefore, an alternative approach is to offload activations to the CPU, thereby reducing the required amount of ranks. There are two characteristics to support the feasibility of this approach:

\begin{itemize}[label=$\triangleright$,leftmargin=*]
    \item[\textcircled{1}] \textbf{Activation is first-in-last-out}: As shown in Figure~\ref{fig:per_layer_offload}, given any sequence, during the forward propagation, it will be processed sequentially by transformer layers, and activations will be gradually accumulated until reaching a peak after the final layer. Subsequently, during the backward propagation, these activations will be consumed from the last layer to the first one. Since the activations produced by earlier layers are used more later (i.e. FILO), it is promising to offload these activations to the CPU during the forward propagation and reload them back into GPU when needed in the backward propagation.
    \item[\textcircled{2}] \textbf{$O(N^2)$ computation can overlap $O(N)$ offloading}: It is well-known that transferring data between GPU and CPU is typically inefficient due to the limited PCIe bandwidth. The offloading time usually far exceeds the computation time, making it impractical. Fortunately, as mentioned in §\ref{sec:background_long_context_training}, the computational complexity of attention is $O(S^2)$, while the memory complexity is $O(S)$. Therefore, for sufficiently long sequences, the $O(S^2)$ computation time will inevitably surpass the $O(S)$ data transfer time, allowing the offloading to be perfectly masked under computation.
\end{itemize}

As illustrated in Figure~\ref{fig:offload}(b), we designed a general component named \textit{act\_ctx} (Listing~\ref{lst:act_ctx}) to support activation offloading. This component maintains two cuda streams for D2H (Device-to-Host) and H2D (Host-to-Device) separately. It automatically captures activation tensors from the computation graph and offloads them to the CPU (use \textit{asyncCudaMemcpy} API) at appropriate times during the forward propagation, and establishes asynchronous dependencies between the D2H stream and the computation stream. The original tensor in the computation graph is replaced with the metadata $\{ \text{layer id}, \text{act id} \}$. Similarly, during the backward propagation, the metadata stored in the computation graph is used to index and reload corresponding activations in the H2D stream. Figure~\ref{fig:per_layer_offload} illustrates the whole process. The \textit{act\_ctx} also supports a parameter named \textit{offload\_ratio}, providing token-level fine-grained control over the proportion of activations offloaded to the CPU. This capability balances GPU memory savings with optimal overlap of computation.

% \begin{figure}[!t]
\begin{lstlisting}[
    caption={usage of \textit{act\_ctx}}, 
    label={lst:act_ctx},
    abovecaptionskip=2pt,
    belowcaptionskip=2pt
]
# Separate offload_ratio to each micro-batch
act_ctx = get_act_ctx(num_micro_batch, offload_ratios)
# forward of micro-batch-i
act_ctx.update_micro_batch_id(i)
with act_ctx:
    forward_func(...)
# backward of micro-batch-j
act_ctx.update_micro_batch_id(j)
with act_ctx:
    backward_func(...)
\end{lstlisting}
% \vspace{-22pt}
% \end{figure}

\mysubsubsection{Selective Offloading}
Activation offloading leverages CPU memory to alleviate the burden on GPU memory. However, only for long sequences the computation can perfectly overlap with offloading. This means we cannot offload all tokens assigned to each rank indiscriminately. Instead, we must selectively offload each token based on the FLOPs. 

Assume the number of layers per rank as $l$, the token capacity per rank as $C$. Given a sequence with length $s_i \ge C$, we define the computation time and activation size for each layer as $T(s_i)$ and $\text{Act}(s_i)$, respectively. The bandwidths of D2H and H2D are profiled as $B_{\text{d2h}}$ and $B_{\text{h2d}}$. 
We aim to find the offload ratio $r$ that minimizes the required number of HDP ranks $D(s_i)$ for $s_i$ by Eq.~(\ref{eq:offload_ratio}), where $\alpha_1$, $\beta_1$, $\alpha_2$, $\beta_2$ and $\gamma$ are coefficients we profiled for the cost model.

\begin{equation}\label{eq:offload_ratio}
\small
\begin{aligned}
&\argmin_{r} D(s_i), \\
\text{s.t.} \quad T(s_i) &= \alpha_1 s_i^2 + \beta_1 s_i + \gamma,\ \text{Act}(s_i) = \alpha_2 s_i + \beta_2, \\
D(s_i) &= \lceil \frac{2\times \text{Act}(s_i) + (1-r) \times (l-2) \times \text{Act}(s_i)} {l\times \text{Act}(C)} \rceil, \\
T(s_i) &\geq \frac{\text{Act}(s_i) \times r}{\min(B_{\text{d2h}}, B_{\text{h2d}})}, \\
1 &\geq r \geq \min(1, \frac{l \times \text{Act}(C)}{(l-2) \times \text{Act}(s_i)}).
\end{aligned}
\end{equation}

Since different micro-batches have mutual independent forward and backward propagation, in Listing~\ref{lst:act_ctx} we assign a separate \textit{offload\_ratio} derived from Eq.~(\ref{eq:offload_ratio}) to each micro-batch. This method effectively compresses the number of ranks required for long sequences from $\frac{s_i}{C}$ to $D(s_i)$, as shown in Figure~\ref{fig:offload}(a). It not only significantly reduces communication overhead but also enables the more available HDP ranks to process data, thereby improving efficiency. 

% \begin{figure}[!t]
\begin{algorithm}[!t]
% \small
\caption{Naive HDP Solution}
\label{alg:naive_hdp}
% \KwIn{Global Batch $\mathbb{B} = \{s_1, s_2, \ldots, s_n\}$, HDP Rank Capacity $C$}
\adjustbox{minipage={1.1\linewidth}, margin=0pt 2pt}{\KwIn{%
    Global~Batch $\mathbb{B} \!=\! \{s_1, s_2, \ldots, s_n\}$, Rank~Capacity $C$
}}
\For{each sequence $s_i \in \mathbb{B}$}{
    Determine offload ratio $r$ and minimum required number of HDP ranks $D(s_i)$ using Eq.(\ref{eq:offload_ratio})\;
    \If{$d_i == 0$}{
        Add $s_i$ to $pack\_list$\;
    }\Else{
        Update $map^r[s_i] \gets r$ and $map^d[s_i] \gets D(s_i)$\;
    }
}
\While{$pack\_list$ is not empty}{
    Pack $subset$ by best-fit strategy to fill capacity $C$\;
    Update $map^r[subset] \gets 0$, $map^d[subset] \gets 1$\;
}
Assign sequences to $d_{hdp}$ HDP ranks based on $map^d$\;
Initialize $act\_ctx$ for each micro-batch using $map^r$\;
\textbf{Return} micro-batches, $act\_ctx$ for each HDP rank
\end{algorithm}
% \end{figure}

\begin{figure*}[!t]
\centering
\includegraphics[width=\linewidth]{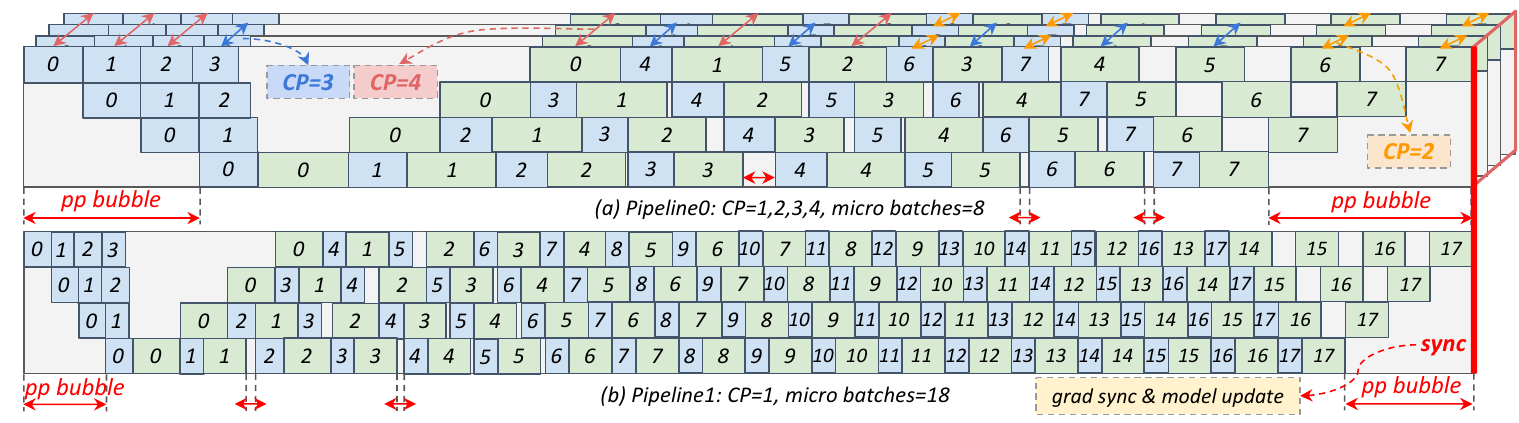}
\vspace{-20pt}
\caption{\small{Balanced Data and Pipeline Parallelism}}
\label{fig:dp_pp_balance}
\vspace{-10pt}
\end{figure*}
% \vspace{-10pt}

\mysubsubsection{Overlap Efficiency Discussion}
As we know, the NCCL communication needs to occupy a portion of streaming multiprocessors (SMs), to reach the peak bandwidth over InfiniBand and NVLink. Consequently, even with communication-computation overlap, the computation kernels cannot fully utilize all the tensor cores, resulting in inefficiencies. Fortunately, the D2H and H2D kernel use the DMA engine rather than SMs, making it overlap perfectly with both computation and communication. Moreover, we use cached pinned host memory to further reduce the overhead of CPU memory allocation and speed up the data exchange between the device and host. Since pipeline parallelism interleaves the forward and backward propagation of different micro-batches, the D2H and H2D kernels could execute simultaneously, thereby maximizing the bidirectional bandwidth of PCIe.

\subsection{Overall Routine}
The overall routine of \system is outlined in Alg.~\ref{alg:naive_hdp}. Briefly speaking, the algorithm traverses each sequence $s_i$ in the global batch. For long sequences, it derives the offload ratio $r$ and determines the required number of ranks $D(s_i)$ (lines 1-6). For short sequences, it packs them to fill each rank's capacity $C$ (lines 7-9). The processed sequences are then assigned to $d_{\text{hdp}}$ ranks, and the algorithm returns the micro-batches and $act\_ctx$, for execution (lines 10-12).

\section{Balance Scheduler}
\label{sec:method_balance_scheduler}

In this section, we introduce the balance scheduler to address both the DP and PP imbalance issues. By carefully orchestrating data assignment (instead of line 10 in Alg.~\ref{alg:naive_hdp}), it mitigates these imbalances while keeping the minimum communication as §\ref{sec:method_comminication_optimizer} performs. We will first outline several key insights and then propose our heuristic solution.

\subsection{Redefine micro-batch}
\begin{figure}
\centering
\includegraphics[width=\linewidth]{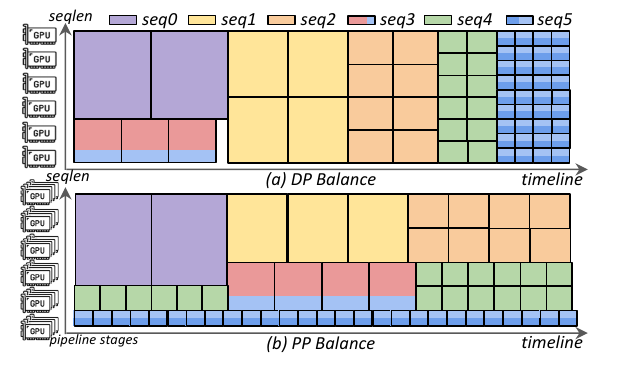}
\vspace{-20pt}
\caption{\small{Balance Strategy}}
\label{fig:balance_strategy}
\vspace{-10pt}
\end{figure}

Gradient accumulation requires that different DP ranks execute the same number of micro-batches, based on the assumption that all micro-batches have the same computational load. However, as mentioned in §\ref{sec:motivation_imbalanced_computation}, execution times for different micro-batches can significantly vary. In \system, we redefine a more flexible strategy, which enables different HDP ranks to process different numbers of micro-batches (same size but differ in workloads), to mitigate the imbalance issue. As shown in Figure~\ref{fig:balance_strategy}, it makes all the ranks finish computation at the same time. More importantly, this strategy does not affect model convergence. Regardless of how sequences are assigned to HDP ranks, we finally calculate the sum of gradients from all tokens in the global batch, as discussed in §\ref{sec:method_data_aware_sharding_and_communication}, which ensures the mathematical equivalence.

\begin{algorithm}[!t]
% \small
\SetAlgoLined
\caption{Balance Strategy for HDP}
\label{alg:balance_strategy}
\KwIn{Global Batch $\mathbb{B} = \{s_0, s_1, \ldots, s_n\}$, Rank Capacity $C$, HDP Degree $d_{hdp}$, Delta $\delta$}
\KwOut{micro-batches for HDP Ranks}

\BlankLine
Initialize $\text{micro\_batches} = [\ ] \times d_{hdp}$\;
Initialize $\text{exec\_times} = [0] \times d_{hdp}$\;

\BlankLine
\textbf{\# Step 1: Sort and Bucketize} \\
Sort $\mathbb{B}$ by sequence length in descending order\;
Divide $\mathbb{B}$ into buckets such that each bucket has an approximately equal sum of FLOPs\;

% \BlankLine
\While{buckets is not empty}{
    \textbf{\# Step 2: Identify Target Ranks} \\
    Calculate $\text{max\_time} = \max(\text{exec\_times})$\;
    Determine $\text{target\_ranks} = \{i \mid \text{max\_time} - \text{exec\_times}[i] > \delta\}$\;

    % \BlankLine
    \textbf{\# Step 3: Assign Sequences} \\
    \While{Exist ($\text{max\_time} - \text{exec\_times}[i] > \delta$)} {
        \If{using DP-Balance strategy}{
            Select $seqs$ from the \underline{first bucket}\;   
        } \ElseIf{using PP-Balance strategy}{
            Select $seqs$ sequentially from \underline{all buckets}\;
        }
        Assign $seqs$ to target\_ranks\;
        Update micro\_batches and exec\_times\; 
        Update target\_ranks based on exec\_times\;
        \If{exist bucket is empty}{
            Remove bucket from buckets\;
        }
    }    
}
\BlankLine
\textbf{Return} micro\_batches
\end{algorithm}

\subsection{Solve PP Imbalance}
\label{sec:method_solve_pp_imbalance}

\textit{Insight 1: PP bubbles are less when sequences of different length levels are assigned to separate pipelines.} 

It is crucial to ensure that the pipeline processes micro-batches with similar execution times. As illustrated in Figure~\ref{fig:balance_strategy}(b), when $d_{pp}=4$, any 4 consecutive micro-batches on the timeline will be executed by 4 PP stages at the same time. If their execution times differ significantly, extra PP bubbles occur. Due to the limited number of long sequences in the global batch, some pipelines have to be assigned sequences of multiple length levels. Fortunately, only during transition phases (e.g., when 4 consecutive micro-batches belong to different length levels) will cause extra PP bubbles.

We assign more micro-batches to those pipelines with less average execution times. As illustrated in Figure~\ref{fig:dp_pp_balance}(a)-(b), pipeline-0 handles micro-batches with larger average execution times and is therefore assigned only 8 micro-batches. In contrast, pipeline-1 is assigned 18 micro-batches to synchronize with pipeline-0. Additionally, due to more micro-batches, the bubble rate is further reduced.

\subsection{Solve DP Imbalance}
\label{sec:method_solve_dp_imbalance}

\textit{Insight 2: It is only necessary to maintain load balance at each time step when pipeline parallelism is not applied.}

If only apply DP without PP, achieving load balance only requires that, at any given time, the micro-batches executed by different HDP ranks have similar execution times. There is no need to consider the workload imbalance between micro-batches across different time steps on the timeline.

A straightforward method is to assign sequences of the same length level across different HDP ranks at the same time, as shown in Figure~\ref{fig:balance_strategy}(a). Moreover, we still assign more micro-batches to those ranks that process shorter sequences than others at the same time. Finally, it ensures that all HDP ranks synchronize gradients nearly simultaneously. 

\subsection{Balance Strategy}
Alg.~\ref{alg:balance_strategy} describes the balance strategy. Firstly, we sort the sequences in the global batch $\mathbb{B}$ by length in descending order. These ordered sequences are then divided into buckets with approximately equal sum of FLOPs, and thus the buckets with longer average lengths contain fewer sequences (lines 3-5). Secondly, we determine those ranks that have shorter execution times for later assignments (lines 7-9). Thirdly, if using the DP-Balance strategy, we select sequences from the same bucket. Otherwise, if using the PP-Balance strategy, we select the sequences sequentially from all buckets. In practice, ranks with shorter execution times are assigned with more sequences (lines 12-15). Finally, we repeat the second and third steps until all the buckets are empty. 

\section{Implementation}
\label{sec:method_implementation}
\system is implemented in approximately 16K lines of code based on Python, C++, and CUDA. It has been integrated with MegaScale~\cite{megascale}, a high-performance framework for LLM training. To support large-scale training and communication, we also apply the following optimizations.

\mysubsubsection{GQA}
Group Query Attention (GQA) has become an indispensable feature in modern LLMs (e.g. LlaMA3 and Mistral), it helps reduce the number of KV heads, thereby decreasing the communication volume for dist-attn. All systems mentioned in this paper apply the GQA technique.

\begin{figure}
\centering
\includegraphics[width=\linewidth]{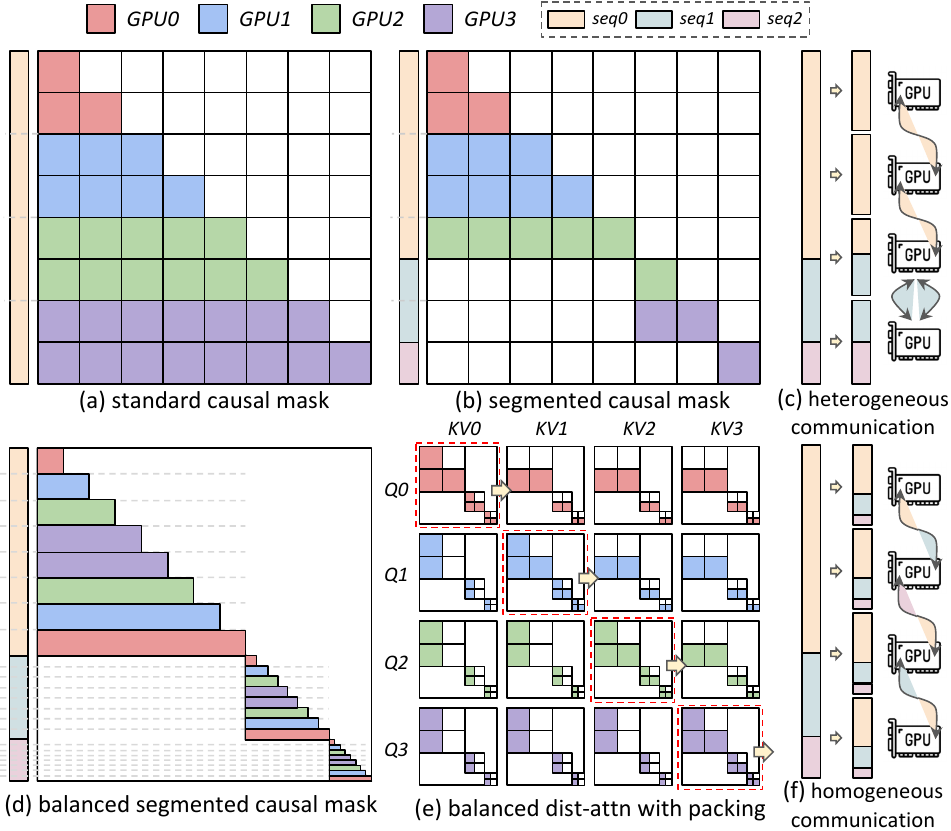}
\vspace{-20pt}
\caption{\small{Dist-attn Optimized for Packed Sequences}}
\label{fig:dist_attn_with_packing}
\vspace{-10pt}
\end{figure}

\mysubsubsection{Dist-attn with Packing}
As workload is proportional to the area of the attention mask, sequentially partitioning the sequence across devices causes workload imbalance. Several techniques~\cite{lightseq,striped_attn,context_parallel} have been proposed to solve this issue. However, they are not suitable for the special segmented causal attention mask for packed sequences. As illustrated in Figure~\ref{fig:dist_attn_with_packing}, to avoid heterogeneous computation and communication within the CP group, we optimize the current dist-attn. Each subsequence of the packed sequences is uniformly divided into $2N$ parts and symmetrically assigned to the $N$ devices. It ensures that each device holds $\frac{1}{N}$ of all the subsequences and covers $\frac{1}{N}$ of the attention mask area. All devices participate in the same ring-P2P communication, with the same data exchange volume. 

% \vspace{-10pt}

\begin{figure}
\centering
\includegraphics[width=\linewidth]{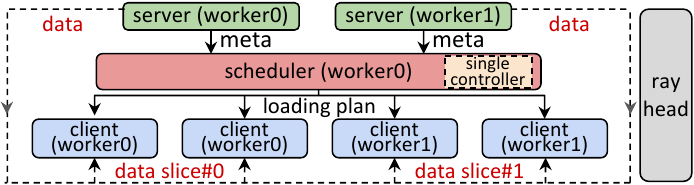}
\vspace{-20pt}
\caption{\small{Remote Dataloader}}
\label{fig:remote_dataloader}
\vspace{-12pt}
\end{figure}
% \vspace{-10pt}

\begin{figure}
\centering
\includegraphics[width=\linewidth]{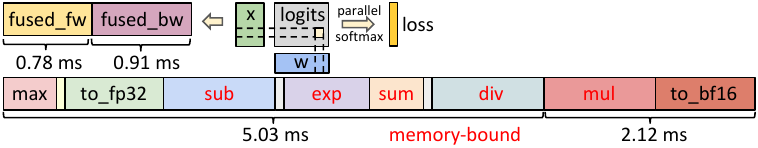}
\vspace{-20pt}
\caption{\small{Fused SoftmaxCrossEntropy}}
\label{fig:fused_ce}
\vspace{-12pt}
\end{figure}
% \vspace{-10pt}

\mysubsubsection{Remote Dataloader}
\system requires global batch information at each training step to schedule data assignment. However, existing dataloader solutions typically follow SPMD (Single Program, Multiple Data) mode, where each rank reads only partial data of the batch. To maintain the global information, all HDP ranks have to read the entire global batch simultaneously, which imposes significant pressure on both network communication and CPU memory. To address this issue, we implement a remote dataloader using Ray~\cite{ray}, which provides real-time scheduling and planning capabilities in a global view. As shown in Figure~\ref{fig:remote_dataloader}, consider a setup with two GPU nodes, worker-0 and worker-1, and one CPU node as the Ray head, there exist three types of roles encapsulated by ray actors. The Server Roles are CPU processes in worker nodes, which fetch and preprocess raw data from HDFS and generate metadata. The Scheduler Role, as the single controller, is a CPU process in worker-0, which collects the global metadata from all servers, deduces the loading plan, and broadcasts it to clients. The Client Roles are GPU processes in worker nodes, which read the partial data from servers based on the loading plan.

\mysubsubsection{Fused SoftmaxCrossEntropy}
Modern LLMs typically use tokenizers with a large vocabulary size (e.g. 128K in LLaMA3~\cite{llama3}, 130K in Mistral~\cite{mistral_tokenizer} and over 150K in Qwen2.5~\cite{qwen2.5}). To stabilize precision, current methods (e.g. VocabParallel in Megatron-LM) convert the logits variable from BF16 to FP32 before calculating the SoftmaxCrossEntropyLoss. However, FP32 logits consume significant memory. For instance, with a context length of 256K and a vocabulary size of 128K, it requires 16GB under TP=8. Besides, the kernels are memory-bound and inefficient. As illustrated in Figure~\ref{fig:fused_ce}, we develop FusedSoftmaxCrossEntropy, which fuses numerous operations into a single kernel, takes BF16 inputs, and still performs online computations in FP32 precision. It saves both time and memory compared to existing methods.
% \clearpage
\section{Experiments}
\label{sec:expr}

\subsection{Experimental Setup}
\label{sec:exp_setup}

\begin{figure*}
\centering
\includegraphics[width=\linewidth]{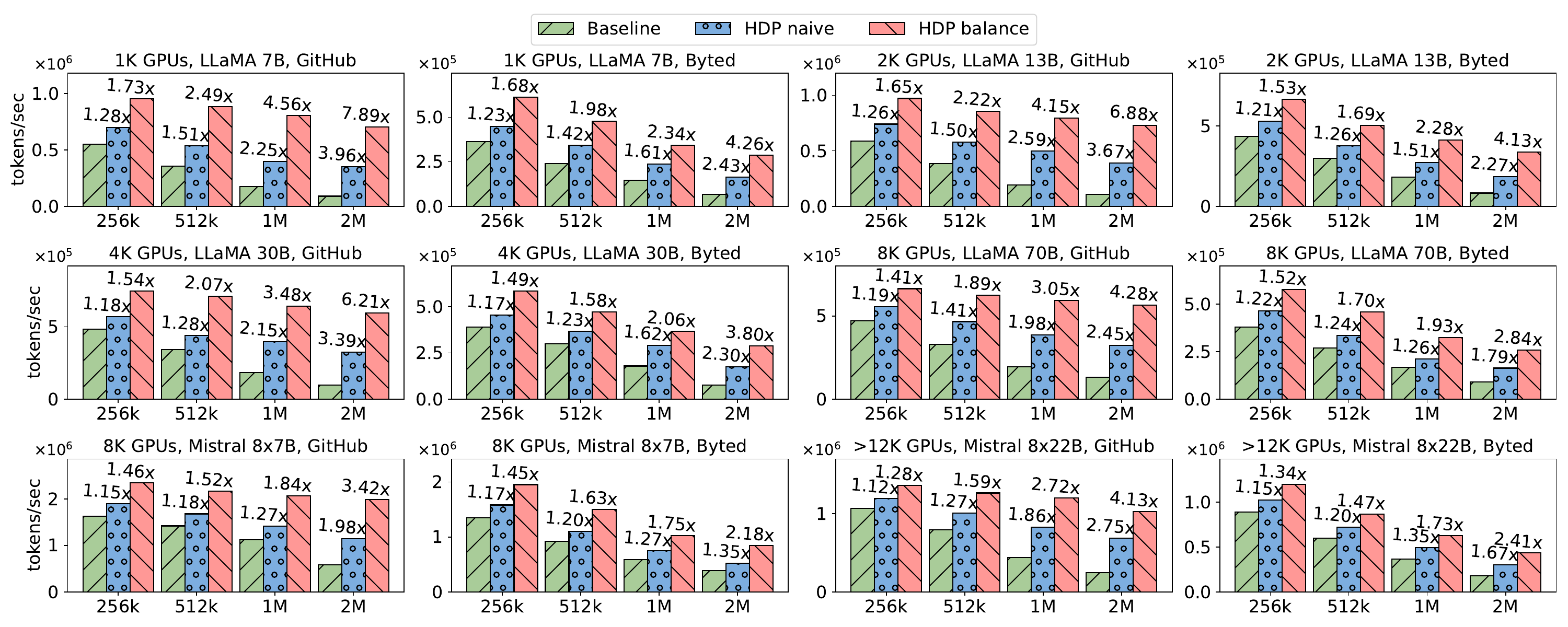}
\vspace{-20pt}
\caption{\small{End-to-end evaluation (measured in tokens per second).}} 
\label{fig:expr_e2e}
\vspace{-5pt}
\end{figure*}

\begin{figure*}
\centering
\includegraphics[width=\linewidth]{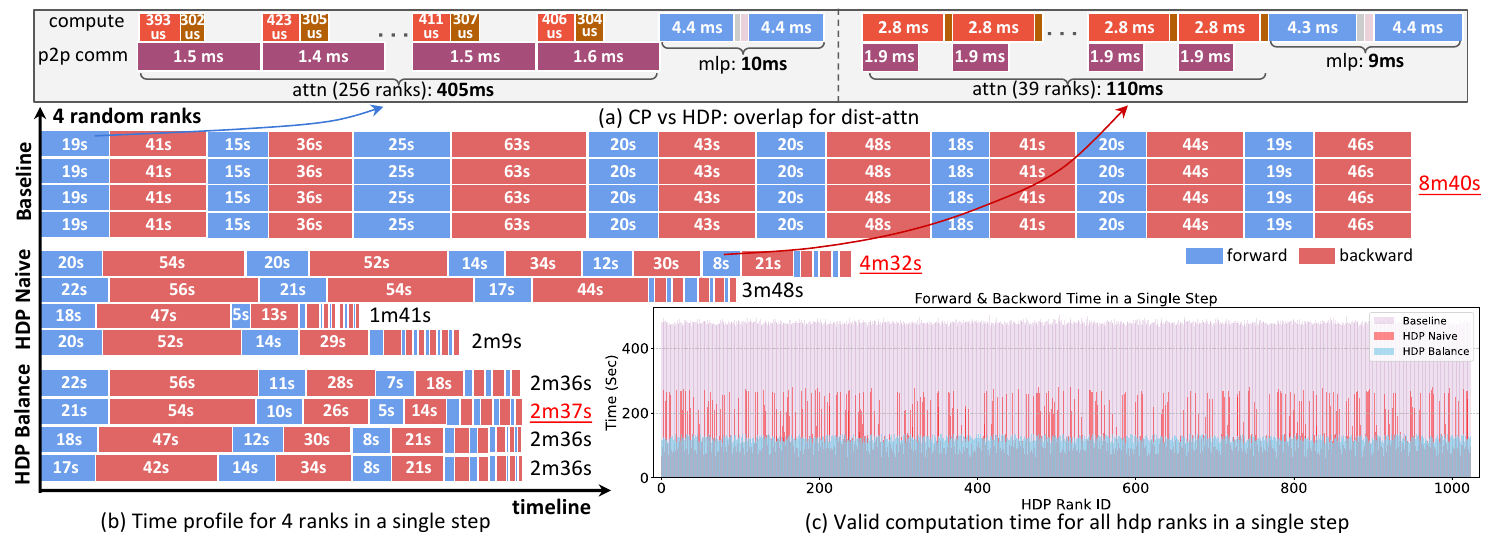}
\vspace{-20pt}
\caption{\small{Case Study}} 
\label{fig:case_study}
\vspace{-10pt}
\end{figure*}

\mysubsubsection{Environments}
% Our experiments are conducted on a large-scale productive GPU cluster with 1024 servers, where each server is equipped with 1800GB memory, 186 CPU cores, and 16 NVIDIA L20 GPUs, total \textbf{16384} GPUs. Within each server, there are 4 PCIE switches, each switch links with 4 GPUs and 2 NICs. The GPUs are connected via PCIE with a bidirectional bandwidth of 64GB/s, and the servers are connected by 8$\times$400Gbps RMDA network.
% based on RoCEv2 with rail-optimized topology.
Our experiments are conducted on a large-scale productive GPU cluster with more than 12,000 GPUs. (\textit{The specific information regarding the productive cluster, such as the number and type of GPUs, is hidden due to business and confidential concerns.})

\mysubsubsection{Baselines}
Our system is built on MegaScale, a productive LLM training framework for large-scale GPU clusters, which has demonstrated superior performance to DeepSpeed and Megatron-LM. 
Thus, we present the advantages of \system by comparison in three cases: \textcircled{1} MegaScale with static parallelism strategies (DP, TP, PP and CP), along with the dist-attn optimization shown in Figure~\ref{fig:dist_attn_with_packing} for packed sequences; \textcircled{2} MegaScale with naive HDP, as described in Alg.~\ref{alg:naive_hdp}, which only applies communication optimizations; \textcircled{3} MegaScale with balanced HDP, as described in Alg.~\ref{alg:balance_strategy}, which applies optimizations for both communication and balance. To achieve a fair comparison, we set the same $d_{tp}$ and $d_{pp}$ for all three cases and set the $d_{hdp}$ in \textcircled{2}\textcircled{3} equal to $d_{dp}\times d_{cp}$ in \textcircled{1}, where the $d_{cp}$ corresponds the minimum required number of ranks to support the context length of model.

\begin{table}[!t]
\small
\centering
\caption{\small{Models for evaluation.}}
\vspace{-10pt}
\label{tb:models}
\begin{tabular}{|c|c|c|c|c|}
\hline
Model & \#Layers & \#Heads & \#Groups & Hidden Dim
\\
\hline
\hline
LLaMA-7B & 32 & 32 & 8 & 4096
\\
\hline
LLaMA-13B & 40 & 40 & 8 & 5120
\\
\hline
LLaMA-30B & 60 & 56 & 8 & 6656
\\
\hline
LLaMA-70B & 80 & 64 & 8 & 8192
\\
\hline
Mistral-8$\times$7B & 32 & 32 & 8 & 4096 (topk=2)
\\
\hline
Mistral-8$\times$22B & 56 & 48 & 8 & 6144 (topk=2)
\\
\hline
\end{tabular}
\vspace{-10pt}
\end{table}

\mysubsubsection{Models and Datasets}
We evaluate our work with both dense and sparse LLMs, as detailed in Table~\ref{tb:models}. For the dense model, we choose the LLaMA-series LLMs with four different sizes, LLaMA-7B, LLaMA-13B, LLaMA-30B, and LLaMA-70B. For the sparse model, we choose the Mistral-series LLMs (MoE) with two different sizes, Mistral-8x7B (active parameters = 13B/47B) and Mistral-8$\times$22B (active parameters = 39B/141B). Two datasets are used in our experiments, i.e., \textit{GitHub} and \textit{Byted}, as we have introduced in §\ref{sec:motivation_data_heterogeneity}. Figure~\ref{fig:sample_token_ratio} illustrates the data distribution for these two datasets.

\mysubsubsection{Workloads and Metrics}
For different types and sizes of models, we scale the context length from 256K to 2M, and the cluster size from 1024 GPUs to more than 12,000 GPUs to assess the performance of \system more comprehensively. The global batch for each training step is fixed to 32M tokens, as it's a common practice in large-scale clusters. We use the throughput (tokens per second) as the primary metric to evaluate the performance. All results are averaged over 200 iterations after a 20-iteration warmup.

\subsection{End-to-End Evaluation}
We first assess the end-to-end performance of three methods by measuring the average throughput at each training step, the overall results are shown in Figure~\ref{fig:expr_e2e}. It turns out that both the HDP naive and balance solutions outperform the baseline, achieving a maximum speedup of $7.89\times$. 

\mysubsubsection{Difference in Scalability}
As context length increases, the baseline with static strategies must increase the CP degree to avoid OOM errors. For shorter sequences within the global batch, we have to pack them and apply the dist-attn shown in Figure~\ref{fig:dist_attn_with_packing}, which suffers from inefficient and redundant communication. For instance, only 9.8\% of the tokens in a global batch are longer than 256K for \textit{GitHub} dataset, and scaling the context length from 256K to 2M, we can observe that the throughput of the baseline decreases nearly $2\times$ whenever context length increases $2\times$. In contrast, under the same conditions, the throughput of the HDP naive solution decreases by $1.23\times$ on average and the throughput of the HDP balance solution decreases by only $1.08\times$ on average. The HDP naive solution reduces communication overhead but leaves some ranks idle due to imbalance. Meanwhile, the HDP balance solution eliminates these bubble times and fully releases the performance enabled by flexible and dynamic communication. Consequently, \system outperforms the baseline by up to $7.89\times$ on the \textit{GitHub} dataset.

\mysubsubsection{Difference in Datasets}
The \textit{Byted} dataset contains more long sequences than the \textit{GitHub} dataset, and there exist 37\% of the tokens in a global batch are longer than 256K. As a result, the average throughput and speedup are lower than that on the \textit{GitHub} dataset. However, because \system provides communication optimizations for both long and short sequences, the speedup can still achieve up to $4.26\times$.

\mysubsubsection{Difference in Parallelism Strategies}
Models like LLaMA-7B, 13B and 30B use parallelism strategies including HDP and TP, thus applying the DP-Balance strategy. In contrast, models like LLaMA-70B, Mistral-8$\times$7B and Mistral-8$\times$22B employ HDP, TP and PP, and we apply the PP-Balance strategy. It can be observed that HDP with DP-Balance achieves a higher speedup, compared to the PP-Balance. For instance, with the \textit{GitHub} dataset and a context length of 2M, the speedup of HDP with DP-Balance is between 6.21$\times$-7.89$\times$, while the speedup of HDP with PP-Balance is only between 3.42$\times$-4.28$\times$. As shown in Figure~\ref{fig:balance_strategy}, the DP-Balance strategy only needs to balance computation at each time step, which is easier to achieve than balance computation across all time steps, as required by the PP-Balance strategy.

\begin{figure*}[t]
\centering
\includegraphics[width=\linewidth]{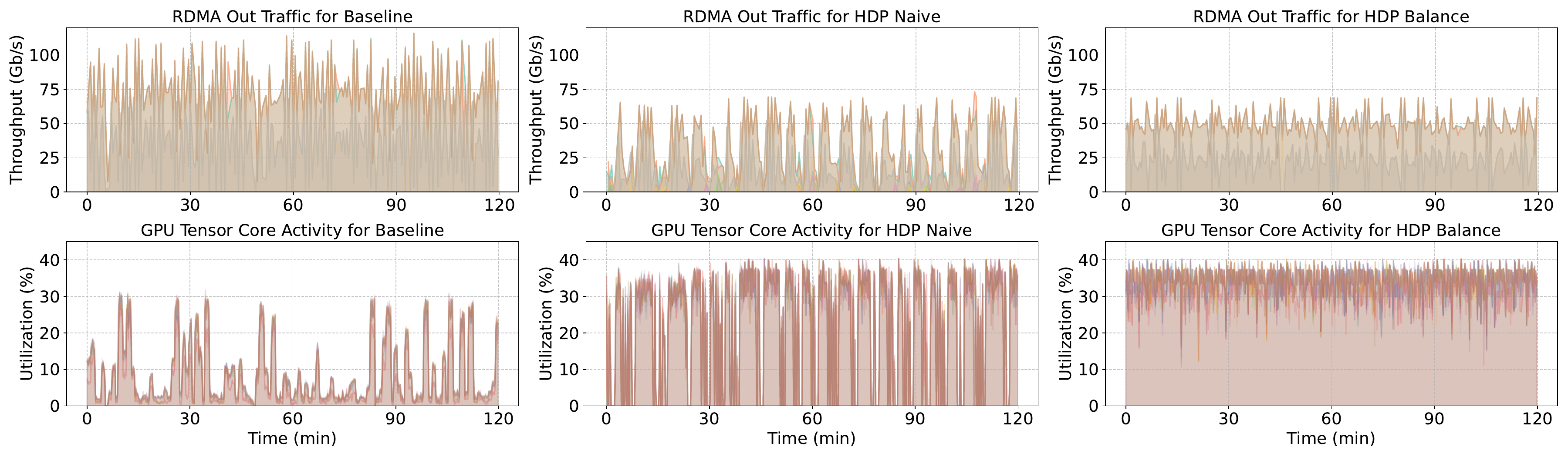}
\vspace{-20pt}
\caption{\small{Network Traffic and Tensor Core Utilization}} 
\label{fig:rdma_gc}
\vspace{-10pt}
\end{figure*}

\begin{figure}
\centering
\includegraphics[width=\linewidth]{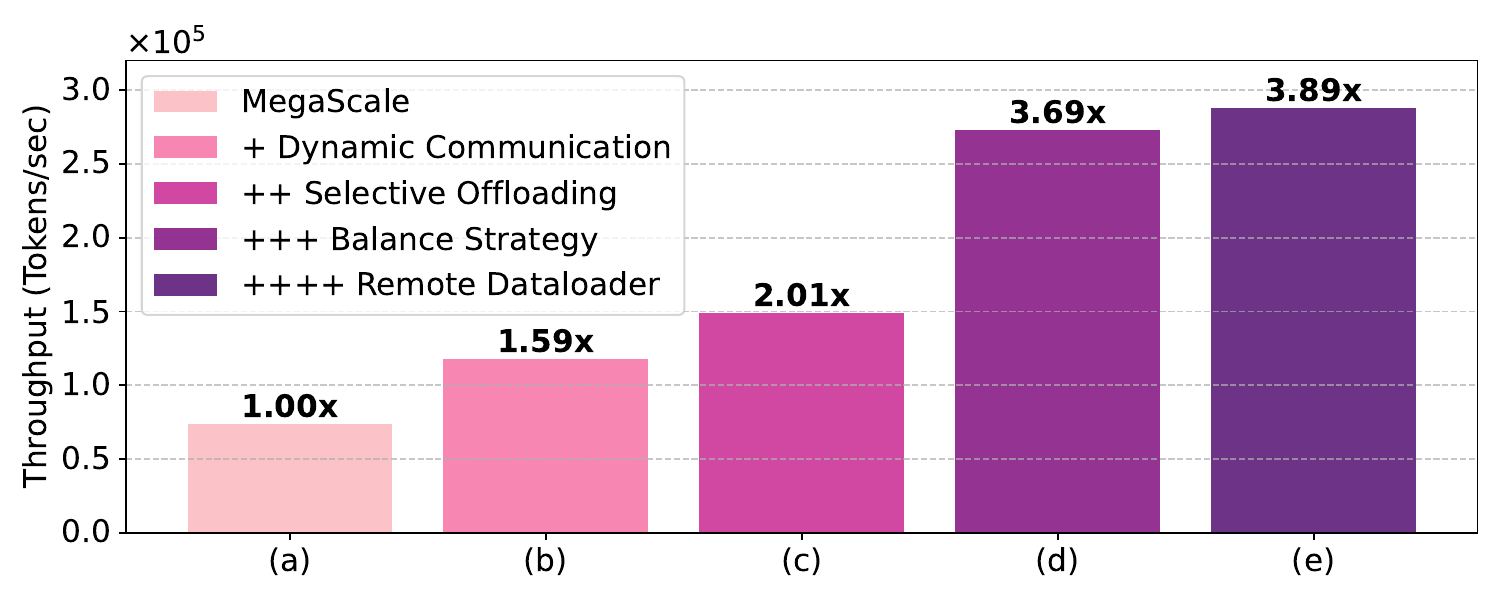}
\vspace{-20pt}
\caption{\small{Ablation Study}}
\label{fig:ablation_study}
\vspace{-10pt}
\end{figure}
% \vspace{-10pt}

\subsection{Case Studies}
\label{sec:expr_case_studies}

To anatomize the super performance of \system more in-deep, we choose the \textit{Byted} dataset and conduct experiments by training LLaMA-7B with 2M context length on a cluster with 1024 GPUs. Figure~\ref{fig:case_study} presents the detailed runtime status of different ranks during a single training step.

\mysubsubsection{Communication-Bound Case}
Firstly, we randomly select 4 ranks from the cluster, and record their forward and backward times within the training step for each method. As illustrated in Figure~\ref{fig:case_study}(b), for the baseline, the number of micro-batches is set as 8, and we have to set $d_{cp}=256$ to support the sequence length of 2M. It can be observed that these 4 ranks exhibit similar execution times. This is because most micro-batches (except the third one) do not have the computational complexity of $O((2M)^2)$, but have to handle the communication volume for 2M. As shown in Figure~\ref{fig:case_study}(a), the P2P communication time far exceeds the computation time, causing the execution time of a micro-batch almost determined by communication (97.6\% of the total time). 

\mysubsubsection{Computation-Imbalance Case}
Under the HDP naive solution, sequences within a global batch are sharded by the minimal required number of ranks. As illustrated in Figure~\ref{fig:case_study}(a), a 312K sequence is sharded by only 39 HDP ranks to serve as micro-batches, and thus the computation time can overlap the communication overhead. However, the training inefficiency still exists due to the imbalance across ranks. As shown in Figure~\ref{fig:case_study}(b), although the third rank completes its 8 micro-batches in 1m41s, it has to wait for the first rank to finish at 4m32s, leading to 171s of idle time. Even so, the HDP naive solution saves 4m8s compared to the baseline.

\mysubsubsection{Balance Case}
Under the HDP balance solution, all ranks finish execution nearly at the same time. As shown in Figure~\ref{fig:case_study}(b), at any time step, each rank is assigned micro-batches with similar FLOPs, and ranks with shorter execution times (e.g. the third and fourth ranks) will be assigned more batches. Consequently, the total time of this step is further reduced to 2m37s, saving 6m3s compared to the baseline.

\mysubsubsection{Overall Comparison}
As shown in Figure~\ref{fig:case_study}(c), we record the valid computation time in a single step for all the 1024 GPUs. It can be found that the HDP naive solution reduces the peak execution time by $1.7\times$ compared to the baseline, but suffers from significant time variance across ranks, with a $4.7\times$ difference between the maximum and the minimum value (min=60s, max=279s, std=68s). The HDP balance solution eliminates the time variance, thereby further reducing execution time by $2.3\times$ compared to the naive solution.

\begin{figure}
\centering
\includegraphics[width=\linewidth]{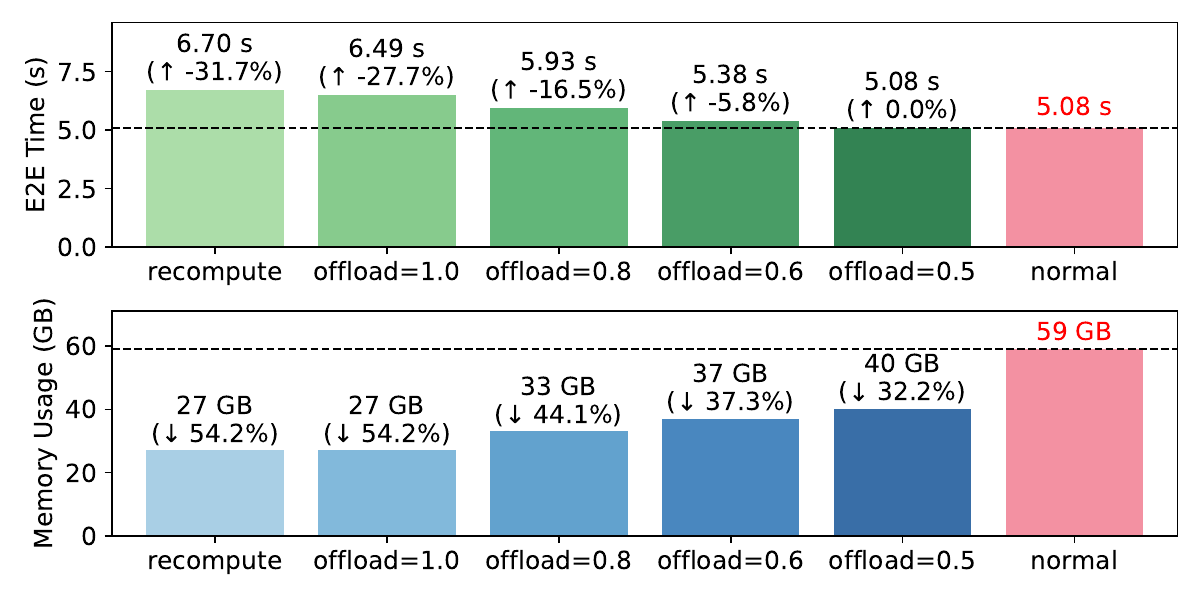}
\vspace{-20pt}
\caption{\small{Effectiveness of Activation Offloading}}
\label{fig:offload_abalation}
\vspace{-10pt}
\end{figure}
% \vspace{-10pt}

\subsection{Ablation Studies}
To dive into the effectiveness of each component within \system, we further conduct ablation experiments using the same configuration as §\ref{sec:expr_case_studies}, as shown in Figure~\ref{fig:ablation_study}. 

\mysubsubsection{Effectiveness of Dynamic Communication}
While Figure~\ref{fig:case_study} provides a snapshot of runtime during a single training step, we further profile the network traffic and tensor core utilization over two hours, as shown in Figure~\ref{fig:rdma_gc}. It can be observed that the baseline exhibits very heavy RDMA traffic, yet the corresponding tensor core utilization is low. This is because most ranks are communication-bound, and the computational units remain idle most of the time due to waiting for redundant communication. This observation is consistent with the situation depicted in Figure~\ref{fig:case_study}(a). When we apply the HDP naive solution, the peak RDMA traffic is nearly halved, indicating that a significant amount of unnecessary communication has been eliminated. Besides, the tensor core utilization also increases from 10\% to 40\%. Thus it achieves a speedup of $1.59\times$ compared to baseline. However, due to the imbalance issue, these improvements are not stable, and both communication and computation hardware units occasionally experience stalls or idle periods.

\mysubsubsection{Effectiveness of Selective Offloading}
Selective offloading serves as a complement to the HDP naive solution. As shown in Figure~\ref{fig:offload_abalation}, activation offloading with ratio $r=1.0$ saves the same memory as recomputation. As the context length is set to 64K, the computation cannot fully overlap with offloading. However, as we reduce the ratio to $r=0.5$, it can save 32.3\% of memory without compromising throughput. Furthermore, the offload ratio is automatically derived from Eq~\ref{eq:offload_ratio}, as the context length increases, a higher ratio can be set to save more memory (e.g. set $r=1.0$ for 256K will not decrease throughput). This method reduces the number of ranks for longer sequences, enabling more sequences to be processed simultaneously with the same number of ranks, thereby improving speedup from $1.59\times$ to $2.01\times$.

\mysubsubsection{Effectiveness of Balance Strategy}
As illustrated in Figure~\ref{fig:rdma_gc}, the balance strategy stables the RDMA traffic and makes the tensor core utilization consistently around 40\%. This indicates that the hardware units of both computation and communication continuously work at full load for over two hours without idling. Consequently, the HDP Balance solution achieves a speedup from $2.01\times$ to $3.69\times$, surpassing the improvements by any other strategy.

\mysubsubsection{Effectiveness of Remote Dataloader}
We employ the remote loader depicted in Figure~\ref{fig:remote_dataloader} and use CPU prefetching to overlap data reading with computation. This approach further improves the speedup from $3.69\times$ to $3.89\times$.

% \section{Related Work}
% \label{sec:related}
% xxx
\section{Conclusion}
\label{sec:conc}
We proposed \system, an efficient, flexible and scalable distributed LLM training framework for large-scale mixed training of long and short sequences. We develop the communication optimizer to eliminate redundant communication and build the balance scheduler to mitigate the imbalanced computation. We evaluate \system on a production cluster with more than 12,000 GPUs, and scale the model size from 7B to 141B and the context length from 256K to 2M, experiment results show that it outperforms MegaScale by up to $7.89\times$.
% \clearpage

\bibliographystyle{ACM-Reference-Format}
\bibliography{reference}

% \onecolumn
% \clearpage
% \appendix
% \input{sections/appendix}

\end{document}